# Development of an Adapter for Analyzing and Protecting Machine Learning Models from Competitive Activity in the Networks Services

Denis Parfenov
Research Institute of Digital Intelligent Technologies
Orenburg State University
Orenburg, Russia
parfenovdi@mail.ru

Anton Parfenov
Research Institute of Digital Intelligent Technologies
Orenburg State University
Orenburg, Russia
anton_parfenov@bk.ru

*Abstract*— **Due to the increasing number of tasks that are solved on remote servers, identifying and classifying traffic is an important task to reduce the load on the server. There are various methods for classifying traffic. This paper discusses machine learning models for solving this problem. However, such ML models are also subject to attacks that affect the classification result of network traffic. To protect models, we proposed a solution based on an autoencoder.**

## I. Introduction

The threat landscape for industrial systems is rapidly evolving, with attacks becoming targeted and motivated. At the same time, machine learning models have also come under attack in recent years. The basis is adversarial attacks that reduce the quality of fine-tuned models. Adversarial attacks are deliberate attempts to trick ML models into making mistakes. They can have serious consequences, such as financial loss, data theft, or even physical damage.

The main problem of existing methods for protecting machine learning models is a priori knowledge about the type of attack carried out on the ML model, as well as the mechanism for carrying it out to identify certain patterns and subsequently counter them. Such information in real conditions cannot always be determined in a timely manner. Due to their vulnerability to malicious attacks, machine learning models used in cybersecurity applications require robust security measures.

The rest of the paper is organized as follows. The second chapter provides an overview of existing literature sources on the topic under study. The third chapter presents the experimental studies of the effectiveness of effectiveness of adversarial attacks on the machine learning models. The fourthchapter contains a conclusion.

## II. Related Work

Scientists around the world are engaged in research in the field of conducting adversarial attacks, assessing the stability of machine learning models in relation to adversarial attacks on machine learning models, as well as developing effective defense methods.

Publication [1] presents the Foolbox Native library in Python for testing the robustness of machine learning models against adversarial attacks. The authors of [2] study the adversarial stability of neural networks from the point of view of robust optimization, showing the high efficiency of the developed solutions on the MNIST data set. In [3], the authors study adversarial attacks in unmanned vehicle systems from the point of view of statistical mechanics and propose a model for interpreting adversarial robustness based on statistical mechanics. Study [4] presents an ensemble strategy for defending against gradient adversarial image recognition attacks based on retraining technique for the CIFAR-10 dataset. Article [5] is devoted to the study of the use of automatic machine learning technology (AutoML) to create complex and secure models, as well as the issues of interpretability of neural network output data to increase resistance to adversarial attacks. Researchers in [6] describe two new approaches to adversarial learning to improve the performance and robustness of a machine learning model. In the first case, the original data sample and its adversarial sample are narrowed in representation space, increasing their distance from different labeled samples, in the second case, the model reconstructs the original data sample from its adversarial representation.

Intelligent forecasting algorithms in 6G communication problems. discussed in [7] within the framework of deep learning methods and their security analysis. The authors proposed a method to mitigate adversarial attacks against proposed 6G machine learning models for millimeter wave (mmWave) beam prediction. The main idea of the research is to obtain erroneous results by manipulating trained deep learning models with a fast gradient sign attack. The proposed model protection concept made it possible to mitigate the adversarial attack on the model, and the mean square errors of the protected model and the unprotected model obtained during the experiments turned out to be very close.

As part of the research into attacks on tabular data, the authors proposed CAPGD, a gradient attack that overcomes the shortcomings of existing gradient attacks using adaptive mechanisms. The authors also proposed an effective evasion attack that combines the CAPGD attack and MOEVA, a better search-based attack. Experimental studies of the listed attacks were carried out on five architectures and four critical use cases [8].

Silvio Russo et al proposed an autoencoder-based anomaly detector that uses unsupervised learning algorithms. The presented approach made it possible to reliably identify various classes of cyber threats without the need for training in specific attacks. The proposed model uses the

The research was funded by the Russian Science Foundation (project No. 22-71-10124).

characteristics of the standard traffic probability distribution for these purposes, which allows detection on unlabeled data [9].

Recent research has shown that deep neural networks (DNNs) are vulnerable to adversarial examples, which can seriously threaten security-sensitive applications. In this paper, we propose an invisible adversarial attack that synthesizes adversarial examples that are visually indistinguishable from benign ones. Two types of adaptive adversarial attacks are proposed: 1) coarse-grained and 2) fine-grained. We conduct extensive experiments on the MNIST, CIFAR-10, and ImageNet datasets, as well as a comprehensive user study with 50 participants [10].

A. Mustapha et al studied DDoS attack detection algorithms using machine learning (ML) and deep learning (DL) algorithms. The research is based on a generative adversarial network (GAN), which was used to generate data suitable for an attack. The authors proposed a DDoS detection method based on the Long Short-Term Memory (LSTM) model. The detection scheme proposed by the authors provides a high level of accuracy in detecting DDoS attacks [11].

Yang Bai et al. study the distribution of adversarial examples for black box attacks. Information about the structure of the image is used as input data. The authors carry out an NP attack using surrogate models. Extensive experiments show that NP-Attack can significantly reduce the number of queries in a black box setup [12].

Fatimah Aloraini proposed a method of adversarial attack using IDS spoofing. The proposed method aims to compromise the IDS by causing it to misclassify attacks. Evaluations of two IDS models—the baseline IDS and the state-of-the-art MTH-IDS—demonstrated significant vulnerability, reducing F1 scores from 95% to 38% and from 97% to 79%, respectively [13].

Thus, a review of modern research in the field of ensuring the security of ML models from various adversarial attacks showed that there is still a lack of effective and practical protection for machine learning models.

This work compares the effectiveness of adversarial attacks on machine learning models in the field of identifying Internet of Things network attacks.

### III. THEORY

Securing machine learning models from attacks is an important area of research because models can be vulnerable to various types of attacks, including confidentiality, integrity, and availability attacks. There are various methods of protection.

Protection against inference attacks:

– Differential Privacy: Add noise to data or model results to protect data privacy.

– Federated Learning: trains a model on distributed data without the need to transfer data to a central repository.

Protection against model extraction attacks:

– Rate Limiting: limiting the number of requests that can be sent to the model in a certain period of time.

– Output Obfuscation: Adds noise to model outputs to make it difficult for an attacker to reconstruct the model.

Protection against adversarial attacks:

– Adversarial Training: Including adversarial examples in the training set to improve the model's resistance to such attacks.

– Defensive Distillation: using knowledge distillation techniques to increase the model's resistance to adversarial attacks.

Protection against data poisoning attacks:

– Data Sanitization: Validating and cleaning data before using it to train a model.

– Robust Training: use of training methods that are resistant to data contamination, such as RANSAC (Random Sample Consensus).

Protecting machine learning models from adversarial attacks in network traffic is an important task, especially in the context of a CICIoT dataset. Adversarial attacks can modify various parameters of network traffic to trick the model into making erroneous predictions. Let's look at what parameters can be changed to attack and how we can protect models from such attacks.

Changing protocols can be used to bypass intrusion detection systems. Use of less common protocols for data transfer. Attackers can also change packet sizes. Packet resizing can be used to mask malicious traffic or create network congestion. Along with the size of the packets, their transmission time and sending frequency may also change. This can create certain patterns that the machine learning model will misclassify. In addition, one of the parameters that can be changed is packet header parameters, such as TTL (Time to Live), DSCP (Differentiated Services Code Point).

### IV. EXPERIMENT

Network traffic classification using machine learning techniques is the process of analyzing and categorizing data passing through a network for various purposes such as anomaly detection, identifying malicious traffic, quality of service management, etc. There are many machine learning methods that can be used for this task. As part of the study, we will use the CIC IoT dataset 2023, which contains traffic data obtained from an IoT topology consisting of 105 devices, 33 attacks are performed.

#### A. Generation of an attack dataset

Generating an attack dataset is a difficult task. In this work, ZooAttack was used, which is a zero-order optimization attack to attack deep neural networks (DNNs). This is an effective black-box attack that only requires access to the inputs and outputs (confidence scores) of the target DNN. The attack uses stochastic descent on zero-order coordinates to directly optimize the target DNN, as well as dimensionality reduction techniques. No transferability or model replacement is required.

Let's look at the basic concepts of ZooAttack. ZooAttack is designed for scenarios where the attacker does not have access to the internal parameters of the model, but only to the outputs of the model, thus simulating a black box attack. The goal of ZooAttack is to create attack examples that look like normal data to a human, but lead to erroneous model predictions. Instead of using gradients, ZooAttack uses gradient estimates based on finite differences. This allows an attacker to estimate gradients by querying the model and analyzing changes in the outputs.

ZooAttack parameters:

– confidence: level of confidence in the attack. The higher the value, the more confidently the model should be wrong;

– targeted: a flag indicating whether the attack is targeted (targeted) or untargeted;

– learning_rate: learning rate for zero order optimization;

– max_iter: maximum number of iterations for optimization;

– binary_search_steps: number of binary search steps to find the optimal value of the constant;

– initial_const: initial value of the constant for binary search;

– abort_early: flag indicating whether optimization should be aborted if no improvement is observed;

– use_resize: a flag indicating whether resizing should be used to speed up the attack;

– use_importance: flag indicating whether feature importance should be used to speed up the attack;

– nb_parallel: number of parallel processes to perform the attack;

– batch_size: batch size for generating adversarial examples;

– variable_h: parameter for estimating gradients using finite differences.

To simulate an attack on a machine learning model for classifying network traffic, the XGBoost model was selected as a shadow model. It is used as a classifier for ZooAttack. To begin with, the XGBoost model was trained on existing data. On the Fig. 1 and 2 show the results of the model metrics on the data generated by the XGBoost model.

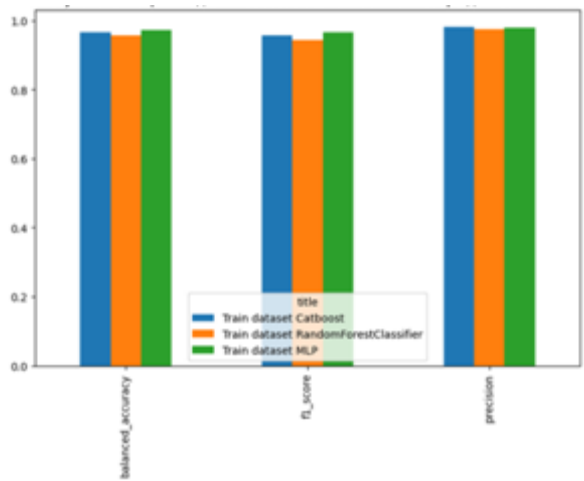

Fig. 1. Metrics on generated data (Train)

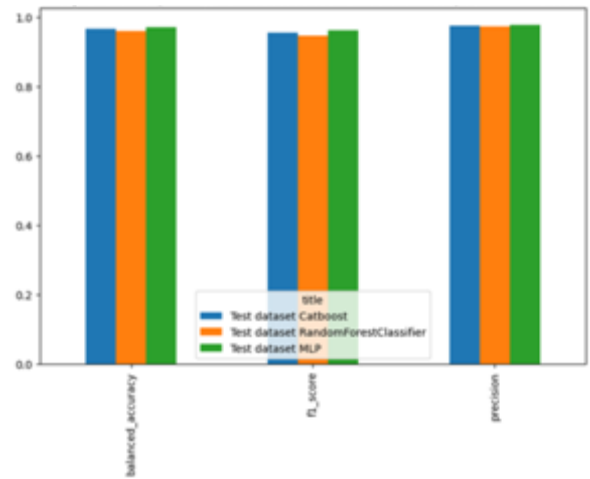

Fig. 2. Metrics on generated data (test)

On the Fig. 3 and 5 show the results of traffic classification models after the ZooAttack attack.

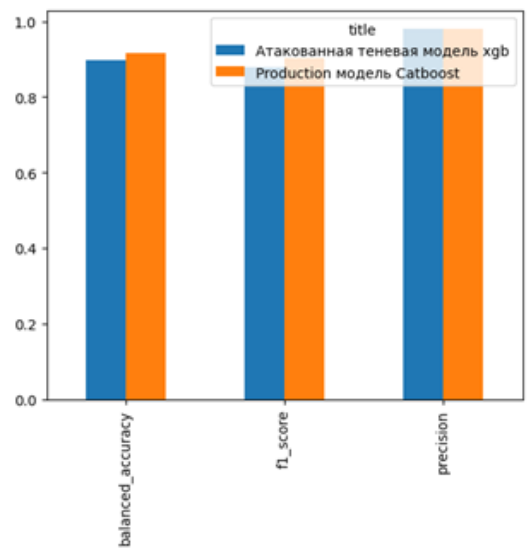

Fig. 3. Attack on the model CatBoost

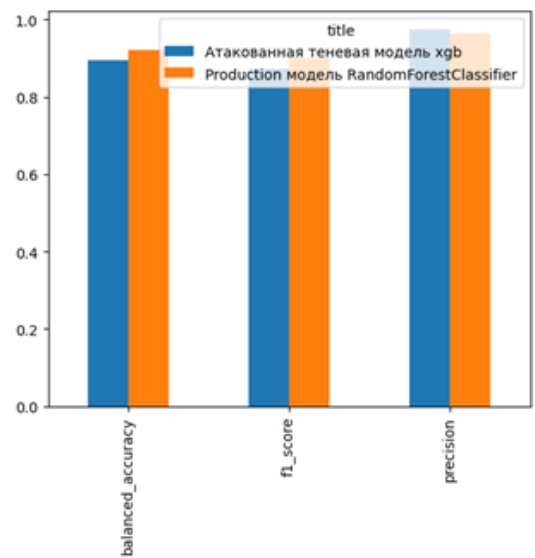

Fig. 4. Attack on the model RandomFosrestClassifier

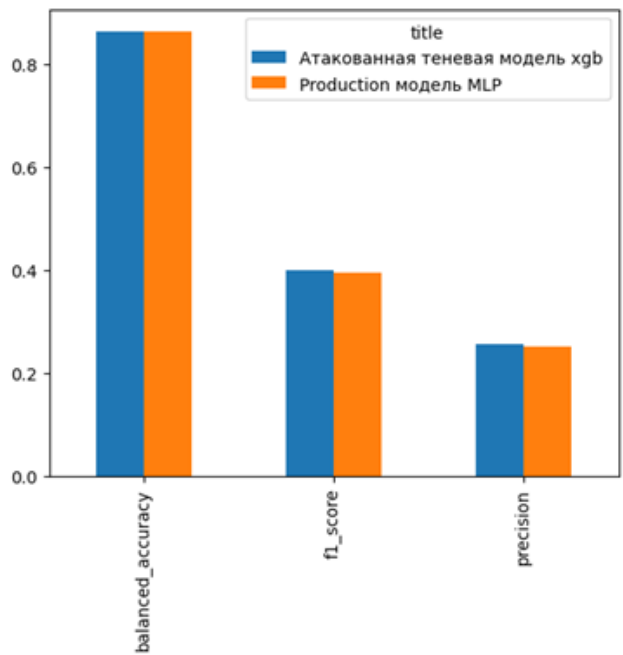

Fig. 5. Attack on the model MLP

*B. Autoencoder protection*

Autoencoders can be used to detect anomalies in data, which helps protect the neural network from external attacks, including data input attacks sent through APIs.

Autoencoder operation algorithm.

1. Training. The autoencoder is trained on a dataset of "normal" traffic.

2. Compression. The autoencoder compresses the input data into a latent representation while preserving the most important information.

3. Recovery. The autoencoder then attempts to reconstruct the original data from the latent representation.

4. Anomaly detection. Data that cannot be recovered well by the autoencoder is considered anomalous.

Due to the fact that traffic on the network contains many parameters, attackers can adjust requests in such a way that this traffic will be classified as normal, although it is attacking. For example, changing the size of the sent packet, the frequency of its sending, connection time. It is also possible to send using inappropriate protocols. Even though the server will return an error, this will increase the load on it.

The autoencoder reduces the number of parameters within itself, and then, restoring their number, displays the original number of parameters. After this, we can compare the original traffic and the one processed by the autoencoder; if the deviation in any parameters exceeds the reference one, which we obtained during training the autoencoder on a dataset that does not contain an attack, a machine learning model that classifies the traffic, then such traffic will be defined as anomalous, i.e. .e containing an attack on the classifier. In addition, the autoencoder allows you to reduce the impact of small changes on the operation of the ML model.

However, this approach increases computational costs when working in real time.

During the work, an autoencoder consisting of 6 layers was implemented. At the input, the autoencoder receives data in which parameters containing timestamps have been removed, as well as information about the IP addresses and ports of the data source and its direction. These settings have been removed because they cannot contain malicious changes. The autoencoder was trained on the CICIoT2023 dataset containing information about the type of traffic in which there were no attacks on the classification model. Next, testing was carried out on the IoT Network Intrusion Dataset (CICIDS2022), containing network traffic generated by 100 IoT devices under attack of 15 different types of attacks, including 5 adversarial attacks.

V. CONCLUSION

Modern attacks aimed at fooling machine learning classifiers in cybersecurity pose a significant threat. Their goal is to distort the operation of algorithms in such a way as to obtain false conclusions or bypass detection of malicious objects. During the research, machine learning models were considered to classify network traffic. These are Random Forest, CatBoost, MLP. These models showed good results in classifying data without attacks, but when attacks were carried out on the data, their accuracy decreased. To reduce the number of incorrect classifications, the dataset was carefully studied and parameters were identified that could be used by attackers to carry out attacks on machine learning models. This is a change in data transfer protocols, a change in the size, transmission time and frequency of packet transmission.

In order to reduce the influence of certain parameters on the result and identify distorted ones, a method of protecting against attacks on machine learning models based on an autoencoder is considered. Thanks to this method, it was possible to identify in advance network traffic containing attack data, mark such data as anomalous, and prevent traffic classifiers from working.

In the future, it will be necessary to continue improving the algorithm, focusing on increasing recognition accuracy, as well as increasing resistance to various data distortions.